\documentclass[]{pasj02} 
\usepackage[switch,mathlines]{lineno} 

\jyear{2025}
\Received{}
\Accepted{}

\begin{document} 

\title{ Energy transfer from jets to surrounding matter to form lateral lobes in SS433/W50 }

\author{
 Hajime \textsc{Inoue},\altaffilmark{1}
 \email{inoue-ha@msc.biglobe.ne.jp} 
}
\altaffiltext{1}{Institute of Space and Astronautical Science, Japan Aerospace Exploration Agency, 3-1-1 Yoshinodai, Chuo-ku, Sagamihara, Kanagawa 252-5210, Japan}




\maketitle

\begin{abstract}
We first investigate an approximate structure of the top region (TR) of a jet, sandwiched by a front shock from which the surrounding matter (SM) inflows and a rear shock from which the jet matter (JM) inflows.
Since pressure in the TR is higher than that in the laterally outer space, both JM and SM flowing in the TR are pressed out from the side of the TR.
Supposing a steady flow of SM and JM there, we construct a simplified two dimensional model on a structure of the TR.
With help of the model, we next infer what happens when precessing jets go through the surroundings in the SS433-W50 system presuming a supernova remnant (SNR) occupies W50.
If we assume reasonable density distributions of the SNR and the interstellar matter in a 10 $\sim$ 100 pc distance range, the density of the surroundings is found to be much higher than that of the jet so that the jet is largely braked in the TR and that outflowing rate of the energy from the side of the TR becomes almost identical to the intrinsic energy flow rate through the jet.
The outflowing energy could spread to the ambient space in a form of a bow shock but the situation of the shock propagation in the present case could be peculiar due to the presence of the precession.
Particularly, all the mass and the energy outflowing from the inner side of the precession cone is considered to be concentrated around the axis of the precession cone. 
As the result, mass-compressed and energy-accumulated regions are expected to appear along the precession axis, which could be the origin of the lobes laterally extending from the main sphere of W50 observed in radio and X-rays.

\end{abstract}


\section{Introduction}\label{Intro}

SS433 is a unique object in steadily exhibiting precessing, mildly relativistic jets (\cite{Margon84}).
The jets have been observed at various distances with respectively different wavebands, 10$^{11} \sim 10^{12}$ cm with X-ray, $\sim 10^{14}$ cm with optical, $\sim 10^{15}$ cm with radio and $\sim 10^{17}$ cm with X-ray again (see \cite{Fabrika04} for the review).
At distance $\gtrsim$ 1 pc from SS433, a spherical structure with radius of $\sim$50 pc accompanied by eastern and western lobes extending to $\sim$ 100 pc from the center are observed around SS433 in radio (e.g. \cite{Downes86}). 
Since the extending directions align to the precession axis, it is currently thought that the elongated structures could be results of interaction between the jets and the surroundings.

Jets had been found from active galactic nuclei (AGN) before the discovery of jets from SS433 and their interactions with the ambient matter were already studied numerically in early 1980s (e.g. \cite{Norman83}). We can see a kind of guideline of understandings by the early works in \citet{Falle91} and relativistic effects on the jet structure in \citet{Marti97}. 

However, a remarkable thing of the SS433-jets is that they exhibit the precession and furthermore the precessing angle apparently reduces from $\sim20^{\circ}$ in the vicinity to the central engine to about 10$^{\circ}$ in the distant lobes.

\citet{Eichler83} proposed a mechanism for a precessing jet to be focused by ambient matter and \citet{Peter93} showed that numerical experiments could really reproduce hydrodynamical collimation on a large scale.
The transition of SS433 jet from a precessing jet with an angle of 20$^{\circ}$ to a continuous hollow non-precessing jet with a smaller opening angle of about 10$^{\circ}$ was confirmed with 3D special relativistic hydrodynamical simulations (\cite{MonceauBaroux14}; \cite{MonceauBaroux15}). Some discussions to support the collimation of precessing jets are given by \citet{Bowler18} 

On the other hand, \citet{Kochanek90} reported that according to their 2D simulations, hollow conical jets, in which jet-matter propagates on the surface of a precessing cone, are prevented from the expansion at some distance.
3D simulations by \citet{Zavala08} also indicated that a precessing jet with precession angle of 20$^{\circ}$ could not reproduce the extending lobes of W50.
Zavala et al. rather proposed a different model where the precessing angle changes with time, starting from zero and growing linearly until it reaches the constant value of 20$^{\circ}$, to well reproduce the expected morphology.
Goodall, \citet{AlouaniBibi10} also obtained from their 3D simulations a conclusion that a history of the jet-activity needs at least three stages to reproduce the lobe-morphology of W50, an early stage of a cylindrical jet without precession, the recent stage of a precessing jet with a precessing angle of 20$^{\circ}$ and a transition stage between the two.

As briefly reviewed above, two models are now presented to reproduce the morphology of W50 extending the eastern and western lobes.
However, no consensus has been obtained on how the lateral lobes have been formed.

In the scenario proposed in this paper, the opening angle of the precession cone is supposed to be kept constant over the entire period of the jet ejection from the beginning up to now.
A simplified two-dimensional model on a top region of a jet formed as the result of interaction between the jet matter and the surrounding matter is studied taking account of matter-outflows from it in direction perpendicular to the jet advancing direction, in section \ref{sec:2}.
The model is applied to the case of the SS433 - W50 system in which the jet direction precesses with the half cone angle of 20$^{\circ}$, in section \ref{sec:3}.
The results indicate that a significant amount of energy outflows from the lateral side of the top region with a rate close to the intrinsic energy flow rate of the jet, which compresses surrounding matter within the precession cone of the jet around the precession axis.
Properties of the compressed region along the precessing axis are semi-quantitatively estimated. They are compared to observations and found to reproduce them fairly well.
Finally, a summary and related discussion are given in section \ref{sec:4}.

The distance to SS433 is assumed to be 5.5 kpc (\cite{Blundel04}; \cite{Lockman07}).

\section{Top region of jets interacting with the surrounding matter}\label{sec:2}
As seen in the above section, understandings of the morphology of the lateral lobes of W50 are still in confusion.
In this paper, we try to analytically understand basic processes governing interactions of jets with surroundings and to have an overall picture to reproduce the lobe morphology.

The structures of the eastern lobe were analytically studied by \citet{Murata96}.
They introduced a simple, one-dimensional model of a structure with three layers of interstellar matter, SNR matter and jet matter, which is formed as the result of collision of the jet to the SNR and interpreted the radio bright region and the X-ray bright region respectively as a region behind the front shock in the interstellar medium and that behind the the reverse shock in the jet medium. 
In this one dimensional model, however, the matter flowing in the shocked regions are premised to accumulate there, which could be unrealistic as discussed below.

\begin{figure}
 \begin{center}
  \includegraphics[width=8cm]{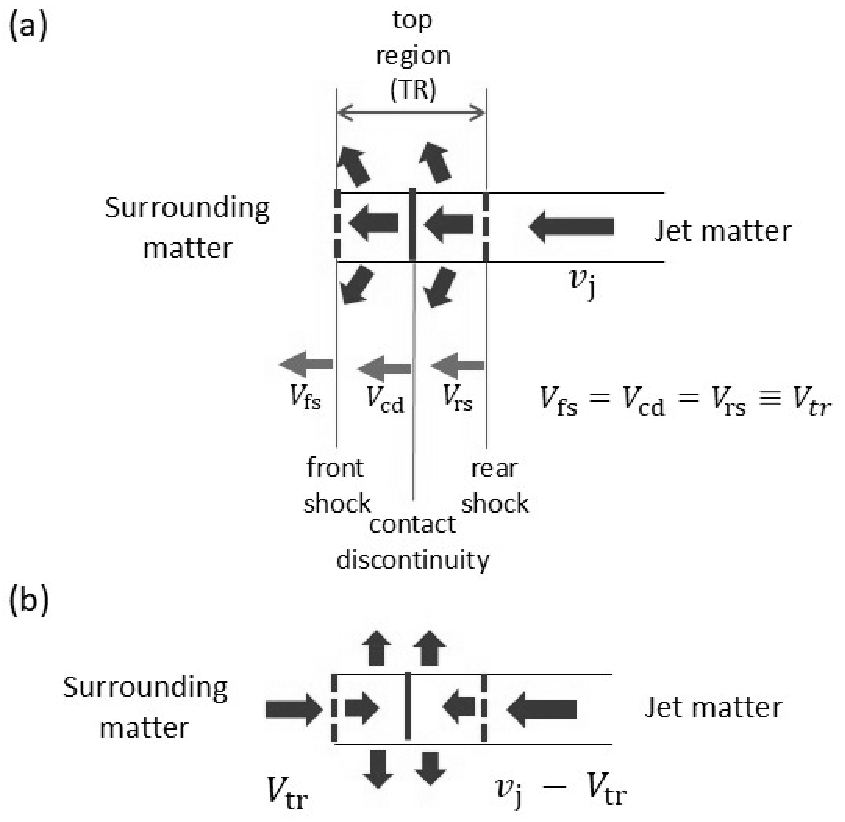} 
 \end{center}
\caption{Schematic cross section of a top region (TR) of a jet in the observational frame (a) and in the TR rest frame (b). {Alt text: two pictures show the basic configuration of the model.}} 
\label{fig:TR}
\end{figure}

Let us consider a situation in which a jet, advancing with a sufficiently supersonic speed and collimated within a cylindrical region, plunges into the surrounding matter (SM) (see figure \ref{fig:TR}-a).
The head of the jet would push the anterior SM, and a front shock should appear at the top of the pushed SM.
On the other hand, the jet matter (JM) pushing the SM would be decelerated  and a rear shock should appear at the backside of the decelerated JM.  
Hereafter, we call the region sandwiched by the front and rear shocks as the top region of the jet (TR).  In the TR, the gas pressure gets as large as the ram pressure of the inflowing matters from the both sides and could be much higher than that of the SM on the lateral side of the TR.  Thus, the matter within the TR is considered to start flowing out in the azimuthal directions perpendicular to the jet advancing axis.
Then, a steady state could be realized in the TR, in that both the SM inflowing from the front shock and the JM from the rear shock eventually flow out from the side of the cylindrical region.
In this case, velocities of the front shock, the rear shock and the contact discontinuity between the SM and the JM should be the same and we represent it with the TR velocity, $V_{\rm tr}$.

Now, we study the structure of the TR in the above situation.
The JM before colliding with the SM is set to run with a velocity, $v_{\rm j}$ in a cone with a half opening angle, $\theta_{\rm j}$ and then the density, $\rho_{\rm j}$, is given as
\begin{equation}
\rho_{\rm j} = \frac{\dot{M}_{\rm j}}{2\pi (r \theta_{\rm j})^{2} v_{\rm j}}
\label{eqn:rho_j}
\end{equation}
at a position with a distance, $r$, from the jet ejection center, 
where $\dot{M}_{\rm j}$ is the total mass flow rate of two opposite jets.
On the other hand, we let $\rho_{\rm sm}$ be the intrinsic density of the SM.
As discussed later, the radial length of the TR can be considered to be well shorter than $r$ on an assumption of $\theta_{\rm j} \ll 1$, and we approximate that the TR is a cylinder with a cross section of $\pi (r \theta_{\rm j})^{2}$.

\subsection{Velocity of the top region} 
In a TR rest frame moving with the velocity of $V_{\rm tr}$ along the jet axis (see figure \ref{fig:TR}-b), the SM flows into from the front shock with the velocity of $V_{\rm tr}$, while the JM does from the rear shock with the velocity of $(v_{\rm j}-V_{\rm tr})$.  Then, the ram pressures from the two sides should balance with each other in the top region, which induces the following equation as
\begin{equation}
\rho_{\rm sm} V_{\rm tr}^{2} = \rho_{\rm j} (v_{\rm j} - V_{\rm tr})^{2} = P_{\rm cd},
\label{eqn:P_cd}
\end{equation}
where $P_{\rm cd}$ is the gas pressure at the both sides of the contact discontinuity.
This equation can be converted to the quadratic equation as
\begin{equation}
x^{2} + \frac{2A}{1-A} x - \frac{A}{1-A} = 0,
\label{eqn:QuadraEq}
\end{equation}
for
\begin{equation}
x = \frac{V_{\rm tr}}{v_{\rm j}},
\label{eqn:x}
\end{equation}
with
\begin{equation}
A = \frac{\rho_{\rm j}}{\rho_{\rm sm}}.
\label{eqn:A}
\end{equation}
The physically appropriate root of this equation is
\begin{equation}
x = \frac{\sqrt{A}}{1+\sqrt{A}}.
\label{eqn:Root-Eq}
\end{equation}
Note that the same equation was already obtained in previous works (e.g. equation (A9) in \cite{Murata96}; equation (1) in \cite{Marti97}).

\subsection{Mass outflow rates from the top region and the radial length}
In the present situation, the SM and JM inflowing into the top region are supposed to both flow out from the side surface steadily.
The outflowing velocities of the SM, $v_{\rm sm, out}$, and the JM, $v_{\rm j, out}$, could be proportional to the respective sound velocities, and are given with the help of equation (\ref{eqn:P_cd}) as
\begin{eqnarray}
v_{\rm sm, out} &=& \beta \sqrt{\frac{5 P_{\rm cd}}{12 \rho_{\rm sm}}} \nonumber \\
&=& \beta \sqrt{\frac{5}{12}} V_{\rm tr}
\label{eqn:v_sm,out}
\end{eqnarray}
for the SM, and 
\begin{eqnarray}
v_{\rm j, out} &=& \beta \sqrt{\frac{5 P_{\rm cd}}{12 \rho_{\rm j}}} \nonumber \\
&=& \beta \sqrt{\frac{5}{12}} (v_{\rm j} - V_{\rm tr})
\label{eqn:v_j,out}
\end{eqnarray}
for the JM.
Here, we have assumed that the specific heat ratio equals 5/3 and that the density on the down stream side of the shock is 4 times as large as that on the up stream side.  $\beta$ is a proportional constant about a few tenth.
The mass-outflow rate from the side of the SM part of the TR could approximately be given as
\begin{equation}
\dot{M}_{\rm sm, out} \simeq 2\pi r \theta_{\rm j} d_{\rm sm} 4 \rho_{\rm sm} v_{\rm sm, out},
\label{eqn:Mdot_sm,out}
\end{equation}
while the mass inflow rate from the front shock is
\begin{equation}
\dot{M}_{\rm sm, in} = \pi (r \theta_{\rm j})^{2} \rho_{\rm sm} V_{\rm tr}.
\label{eqn:Mdot_sm,in}
\end{equation}
$\dot{M}_{\rm sm, in} = \dot{M}_{\rm sm, out}$ should be established in the steady state and thus we obtain with the help of equation (\ref{eqn:v_sm,out})
\begin{equation}
d_{\rm sm} = \frac{1}{4 \beta} \sqrt{\frac{3}{5}} r \theta_{\rm j} .
\label{eqn:d_sm}
\end{equation}
For the JM part of the TR, the mass-outflow rate from the side could be as
\begin{equation}
\dot{M}_{\rm j, out} \simeq 2\pi r \theta_{\rm j} d_{\rm j} 4 \rho_{\rm j} v_{\rm j, out},
\label{eqn:Mdot_j,out}
\end{equation}
and the mass inflow rate from the front shock is
\begin{equation}
\dot{M}_{\rm j, in} = \pi (r \theta_{\rm j})^{2} \rho_{\rm j} (v_{\rm j} - V_{\rm tr}).
\label{eqn:Mdot_j,in}
\end{equation}
$\dot{M}_{\rm j, in} = \dot{M}_{\rm j, out}$ yields with the help of equations (\ref{eqn:x}) and (\ref{eqn:v_j,out})
\begin{equation}
\dot{M}_{\rm j,out} = \frac{\dot{M}_{\rm j}}{2} ( 1 - x ),
\label{eqn:Mdot_j,out-x}
\end{equation}
and 
\begin{equation}
d_{\rm j} = \frac{1}{4 \beta} \sqrt{\frac{3}{5}} r \theta_{\rm j} .
\label{eqn:d_j}
\end{equation}
Thus, considering $\beta \sim$ a few tenth, we can get $d_{\rm sm} = d_{\rm j} \sim r \theta_{\rm j}$.

\subsection{Enegy outflow rate from the top region}
The energy-inflow rate, $\dot{E}_{\rm tr}^{\prime}$, to the TR in the TR rest frame is expressed as
\begin{equation}
\dot{E}_{\rm tr}^{\prime} \simeq \dot{M}_{\rm sm, in} \frac{V_{\rm tr}^{2}}{2} + \dot{M}_{\rm j, in} \frac{(v_{\rm j}-V_{\rm tr})^{2}}{2}.
\label{eqn:Edot_tr^prime}
\end{equation}
In the steady state, this rate can be considered to be the energy-outflow rate from the TR to the surrounding.
In the observational frame, the outflowing matter from the side of the TR should have the velocity component $V_{\rm tr}$ along the jet axis and thus the energy outflow rate from the TR, $\dot{E}_{\rm tr}$ relates to $\dot{E}_{\rm tr}^{\prime}$ as
\begin{eqnarray}
\dot{E}_{\rm tr} &=& \dot{E}_{\rm tr}^{\prime} + \dot{M}_{\rm j, out}\frac{V_{\rm tr}^{2}}{2} + \dot{M}_{\rm sm, out} \frac{V_{\rm tr}^{2}}{2} \nonumber \\
&\simeq& \dot{M}_{\rm sm, in} V_{\rm tr}^{2} + \dot{M}_{\rm j, in} \left[ \frac{(v_{\rm j} - V_{\rm tr})^{2}}{2} + \frac{V_{\rm tr}^{2}}{2} \right] \nonumber \\
&=& \dot{E}_{\rm j} \frac{1}{(1+\sqrt{A})},
\label{eqn:Edot_tr}
\end{eqnarray}
where $\dot{E}_{\rm j}$ is the energy flow rate through one of the two jets given as
\begin{equation}
\dot{E}_{\rm j} = \frac{\dot{M}_{\rm j}}{2} \frac{v_{\rm j}^{2}}{2} ,
\label{eqn:Edot_j}
\end{equation}
and the above expression changes have been done with the help of equations (\ref{eqn:Mdot_sm,in}), 
(\ref{eqn:Mdot_j,in}), (\ref{eqn:rho_j}), (\ref{eqn:x}), (\ref{eqn:A}), (\ref{eqn:Root-Eq}), $M_{\rm sm, in} = M_{\rm sm, out}$ and $M_{\rm j, in} = M_{\rm j, out}$.\\

The above simple model roughly tells us how largely the speed of the jet is decelerated in the TR and how much fraction of the energy flow rate of the jet is scattered to the surrounding is, as a function of $A$ (ratio of $\rho_{\rm j}$ to $\rho_{\rm sm}$).

When $\rho_{\rm j} \gg \rho_{\rm sm}$ ($A \gg 1$), $x$ gets almost unity as seen from equation (\ref{eqn:Root-Eq}), meaning that the deceleration of the jet due to the interaction with the SM is very small. The mass and energy outflowing rates from the TR are also very small as seen from equations (\ref{eqn:Mdot_j,out-x}) and (\ref{eqn:Edot_tr}).

When $\rho_{\rm j} \ll \rho_{\rm sm}$ ($A \ll 1$), on the other hand, $x$ becomes significantly smaller than unity, indicating that the top of the jet is significantly decelerated by the interaction with the SM. The almost all of the mass and energy flow rates through the jet are converted to the mass and energy flow rates from the side of the top region to the surrounding space.

These density dependencies of TR properties are quite reasonable and consistent with results of numerical simulations (see e.g. Figure 16 in \cite{Norman83}).

We can infer fates of matters ejected from TR based on previous relevant works (see e.g. Figure 2 in \cite{Norman83} or Figure 1 in \cite{Falle91}).
There could be four structures induced by the outflows from the TR:
(1) The SM outflowing in the azimuthal directions over 2$\pi$ from the side wall of the TR has a momentum in the jet advancing direction and thus collides with other SM around the jet axis, which forms a shock wave with an umbrella-like shape along the jet axis having the TR at the top, usually called as ``bow shock".
(2) The SM are successively taken in the backside of the bow shock and flows backwards relative to the bow shock. This region is often designated ``Swept up SM region". 
(3) The JM is ejected from the side wall of the TR and flows backwards on the inner side of the swept up SM region. The jet gas continuously ejected from the TR is considered to eventually form a low density region around the jet beam. This region is usually nicknamed as ``Cacoon" but called ``Ejected JM region" here.
(4) The swept up SM and the ejected JM contact with each other at a boundary surface, which is called ``Contact discontinuity".


\section{Jet - surroundings interactions in SS433/W50}\label{sec:3}
Let us apply the above considerations to the case of the jets on the 10 - 100 pc scale in the SS433/W50 system.

\subsection{helical jet-trajectory on the surface of a precession cone}
We suppose a situation in which jet-matter is continuously ejected with a velocity, $v_{\rm j}$, confined in a solid angle of $\pi \theta_{\rm j}^{2}$ around a radial direction from a source and the ejection-direction precesses with a precessing angle of $\theta_{\rm p}$ and a precession period of $t_{\rm p}$.
Because of the precession, the azimuth of the ejection direction at a distance, $r$, changes by $2\pi r \theta_{\rm p} \Delta t/t_{\rm p}$ during a minute time, $\Delta t$, while the radial distance of a jet element varies by $v_{\rm j} \Delta t$.
Hence, the continuously ejected matter makes an apparent trajectory in a form of helix on the surface of the precessing cone with a half opening angle of $\theta_{\rm p}$, having a pitch angle, $\alpha$, given as
\begin{equation}
\alpha = \arctan \frac{d_{\rm gap}}{2\pi r \theta_{\rm p}},
\label{eqn:alpha}
\end{equation}
where 
\begin{equation}
d_{\rm gap} = v_{\rm j} t_{\rm p} = 3.5 \times 10^{-2} \mbox{ pc}
\label{eqn:d_gap}
\end{equation}
is the separation distance in a radial direction between two adjacent paths of the helical structure on the precession cone.
We have adopted $v_{\rm j}$ = 0.26 $c$ ($c$ : the velocity of light) and $t_{\rm p}$ = 162 d. 
In a region of $r \lesssim$ 0.1 pc, the pitch angle of the helix is still large and the helical structure is really observed in radio (e.g. \cite{Blundell04}). The jet behaviors observed in various wavelengths are fiarly well understood (e.g. \cite{Fabrika04} for the review; see also \cite{Inoue22} for discussions on interactions between the coherently precessing jets and disk wind.).

On the other hand, properties of the jets in a region of $r \gtrsim$ 1 pc are still in a fog as briefly introduced in section \ref{Intro}.
Hereafter, we concentrate discussions on the jets in that outer region.

First, we set a premise that the jet matter continuously goes straight there and hence the opening angle of the precession cone on which the jet advances is kept the same value as in the inner region.
Although there is no observational evidence to support this premise, no firm evidence to deny it exits either.

We introduce the polar coordinate ($r, \theta, \phi$) around the precession axis in which $\theta$ is an inclination angle of an $r$ direction from the polar axis and $\phi$ is its azimuthal angle.
In the outer region, the pitch angle of the helical jet is calculated from equation (\ref{eqn:alpha}) as
\begin{equation}
\alpha \simeq 0.9^{\circ} \left(\frac{r}{1\mbox{ pc}}\right)^{-1},
\label{eqn:alpha-Out}
\end{equation}
which is very small.
Since the cross section of the jet trajectory on a plane perpendicular to the radial direction at $r$ is a circle with radius of $r \theta_{\rm j}$, the cross section on a meridian plane with an azimuthal angle of $\phi$ will be an ellipse which has the center at ($r, \theta_{\rm p}, \phi$), the semi-minor axis of $r \theta_{\rm j} \alpha$ in the $r$ direction and the semi-major axis of $r \theta_{\rm j}$ in the $\theta$ direction.
Then, we can have an approximate image of the jet flow along the $r$ direction in which a sheet-like jet element elongating in the plus/minus $\phi$ directions with a radial thickness of
\begin{eqnarray}
d_{\rm r} &\simeq& 2 r \theta_{\rm j} \alpha \simeq \frac{\theta_{\rm j}}{\pi \theta_{\rm p}} d_{\rm gap} \nonumber \\ &\simeq& 1.6 \times 10^{-2} d_{\rm gap}
\label{eqn:d_r}
\end{eqnarray}
and a width in the $\theta$ direction
\begin{eqnarray}
d_{\rm wid} &\simeq& 2 r \theta_{\rm j} \nonumber \\ &\simeq& 3.5 \times 10^{-2} \left(\frac{r}{\mbox{1 pc}}\right)
\label{eqn:d_wid}
\end{eqnarray} 
repeatedly appear in the radial direction with a separation of $d_{\rm gap}$.
In the above numerical estimations, $\theta_{\rm j} = 1^{\circ}$ is assumed.

A notable point is that in the outer region with $r \gtrsim$ 1 pc the width of a jet element, $d_{\rm wid}$, in the direction perpendicular to the flow direction is not less than the separation distance, $d_{\rm gap}$, between the adjacent two in the flow direction, as seen from equations (\ref{eqn:d_gap}) and (\ref{eqn:d_wid}).
The jet elements are considered to run through the ejected JM region surrounding the jet flow. 
The matter in the ejected JM region would tend to enter the gaps between the jet elements and the time necessary for the surrounding matter to fill a gap, $t_{\rm sm, g}$, could roughly be as
\begin{equation}
t_{\rm sm,g} \sim \frac{d_{\rm wid}}{c_{\rm s, sm}},
\label{eqn:t_sm.g}
\end{equation}
where $c_{\rm s, sm}$ is the sound speed of the matter in the ejected JM region.
However, the jet elements passes by in front of the surrounding matter every $t_{\rm j, g}$ defined as
\begin{equation}
t_{\rm j, g} \equiv \frac{d_{\rm gap}}{v_{\rm j}},
\label{eqn:t_j,g}
\end{equation}
and could kick out the surrounding matter even if it enters the gaps.
From above two equations, we see that $t_{\rm j,g}$ should be much shorter than $t_{\rm sm, g}$ since $d_{\rm gap} \lesssim d_{\rm wid}$ as seen above and $v_{\rm j}$ should be much faster than $c_{\rm s, sm}$.
Under the circumstance of $t_{\rm j,g} \ll t_{\rm sm, g}$, there could be almost no room for the surrounding matter to fill the gaps.

It could be possible for matter in the jet elements itself to fill the gaps due to the thermal expansion.
If a pressure of a gap is significantly lower than that of a jet element, the jet element should start thermal expansion. If we approximate for the expansion speed to be a little less than the sound speed of the jet element, $c_{\rm s, je}$, the time scale for the jet element to expand to the size of the gap, $t_{\rm je,e}$ is roughly given as
\begin{equation}
t_{\rm je,e} \sim \frac{d_{\rm gap}}{\beta c_{\rm s, je}},
\label{eqn:t_je,e}
\end{equation}
where $\beta$ is the reduction factor of the expansion speed from the sound speed.
We know that the temperature of the jet elements, $kT_{\rm je}$ ($k$ : the Boltzmann constant), is about 5 keV at $r \sim$ 0.1 pc from the X-ray observation (\cite{Migliari02}).
If we simply assume that the temperature adiabatically cools down due to the density decrease with the $r$ incewase as $\rho_{\rm j} \propto r^{-2}$ associated with the jet motion keeping the conical structure, the temperature decreases as $T_{\rm j} \propto r^{-4/3}$ for the specific heat ratio $\gamma$ = 5/3.
Then, $t_{\rm je,e}$ is expressed as
\begin{equation}
t_{\rm je,e} \simeq 6.4 \times 10^{9} \left(\frac{r}{\mbox{0.1 pc}}\right)^{2/3} \mbox{ s},
\label{eqn:t_je,e-Cal}
\end{equation}
with the help of the relation of $c_{\rm s, j}^{2} = \gamma 2kT_{\rm je}/m_{\rm H}$ ($m_{\rm H}$ : the proton mass) on an assumption of purely hydrogenic gas and $\beta= 0.3$.
On the other hand, the passing time scale for the jet element over the typical distance $r$, $t_{\rm j, r}$, is written as
\begin{equation}
t_{\rm j, r} \simeq \frac{r}{v_{\rm j}} \simeq 4.0 \times 10^{7} \left(\frac{r}{\mbox{0.1 pc}}\right) \mbox{ s}.
\label{eqn:t_j,r}
\end{equation}
The above two time scales are calculated as $t_{\rm je,e} = 6.4 \times 10^{11}$ s and $t_{\rm j,r} = 4.0 \times 10^{10}$ s at $r = 10^{2}$ pc and we see that $t_{\rm je,e}$ is much longer than $t_{\rm j,r}$ even at such a long distance of 100 pc.
This means that the jet passing time over 100 pc is yet insufficient for two adjacent jet-elements to connect with each other due to the thermal expansion.

Hence, we could say that the helical jet structure on the surface of the precession cone with the half opening angle of 20$^{\circ}$ can be maintained even if the TR runs over such a long distance as 100 pc from the central source.

\subsection{Top region of the precessing jets}
In the scenario proposed in this paper, the opening angle of the precession cone is supposed to be kept constant over the entire period of the jet ejection from the beginning up to now.
Then, as discussed in the previous subsection, a long sheet of the jet matter is considered to form a helical structure on the entire surface of the precession cone from the central source to the TR interacting with the surrounding matter.

Since the jet matter in the top one circle of the helical sheet should collide with the surrounding matter and largely be braked there, the following jet matter is expected to run into the front one from behind and merge with it.
As the result, a TR (top region) is expected to be formed as a ring with a radius of $r \theta_{\rm p}$ and a rectangular cross section with the width of $2 r \theta{\rm j}$ and the length of $d_{\rm tr}$. 
Since the typical dynamical time scale of the TR is much longer than the precession period, $t_{\rm p}$, as confirmed below, in spite of the intermittent collision of the following jet matter every $t_{\rm p}$, the model considerations in section \ref{sec:2} can be applied to the present case of TR with some modifications.
First, the density distribution of the jet should be changed as
\begin{equation}
\rho_{\rm j} = \frac{\dot{M}_{\rm j}}{8\pi r^{2} \theta_{\rm p} \theta_{\rm j} v_{\rm j}},
\label{eqn:rho_j-Mod}
\end{equation}
approximating the helical cone structure by the uniform cone one.
Secondly, the expressions in equations (\ref{eqn:Mdot_sm,out}) through (\ref{eqn:d_j}) should also be modified responding to the geometry change of the TR, but the final result of $d_{\rm sm} = d_{\rm j} \sim r \theta_{\rm j}$ does not change.
Then, the dynamical time scale of the TR, $t_{\rm d, tr}$, is roughly estimated as 
\begin{equation}
t_{\rm d, tr} \sim \frac{d_{\rm j}}{V_{\rm tr}} \sim \frac{r \theta_{\rm j}}{V_{\rm tr}} = \frac{r \theta_{\rm j}}{v_{\rm j} t_{\rm p}} \frac{v_{\rm j}}{V_{\rm tr}} t_{\rm p},
\label{eqn:t_d,tr}
\end{equation} 
which is confirmed to be much longer than $t_{\rm p}$, considering $r$ is several 10 pc and $v_{\rm j} \gg V_{\rm tr}$ as calculated later.
Thus, the model in section \ref{sec:2} is assured to be applied to the precession cone configuration of the jet with parameters averaged over one precession cycle and the derivations of $V_{\rm tr}$ in equation (\ref{eqn:Root-Eq}) and of $\dot{E}_{\rm tr}$ in equation (\ref{eqn:Edot_tr}) can be used unchanged.

\subsection{Density distribution of the surrounding matter}
Here, we assume that a supernova remnant (SNR) in the Sedov phase with a radius of $R_{\rm snr} \simeq$ 50 pc exists in W50 and has a shell with a thickness of a tenth $R_{\rm snr}$ and a number density, $n_{\rm snr, sh} = 10^{-1}$ cm$^{-3}$ and the interior with a number density, $n_{\rm snr, in} = n_{\rm snr, sh}/5 = 2 \times 10^{-2}$ cm$^{-3}$.
The value of $n_{\rm snr, sh}$ is selected to roughly reproduce the observed X-ray luminosity from the X-ray brightest region in the eastern lobe as discussed later.
For the interstellar matter (ISM) we adopt a number density, $n_{\rm ism} = n_{\rm snr, sh}/3 = 3.3 \times 10^{-2}$ cm$^{-3}$.
According to \citet{Dickey90}, the average number density of the interstellar H I regions is $\sim 10^{-1}$ cm$^{-3}$ at a position with a height of 200 pc from the Galactic plane, which is the height of SS433 calculated with the Galactic latitude of 2.2$^{\circ}$ and the distance of 5.5 kpc.  This value for HI regions is a little larger than the above value of $n_{\rm ism}$, but it could be acceptable if a part of the interstellar space around the SNR is a thin hot H II region.
Then, the total mass of the ISM swept by the shock wave of the SNR is estimated to be $\sim 430 M_{\odot}$.
(Here and hereafter, the mass densities of the SNR and the ISM are calculated by multiplying the number densities by the proton mass, simply approximating the matter consists of pure hydrogen.)
Since this mass largely exceeds the ejected mass of a few ten $M_{\odot}$ from the super-nova explosion, the density profile of the SNR could be approximated by the Sedov similarity solution (see e.g. \cite{Landau87}) in which the density is expected to jump up to 4 times the ambient density at the shock front and to rapidly decrease toward the explosion center, for the specific-heat ratio of 5/3.
Hence, we assume a simplified density profile for the SNR and ISM as mentioned above.

\subsection{Interactions of the jets with the surroundings}
\begin{figure}
\begin{center}
\includegraphics[width=8cm]{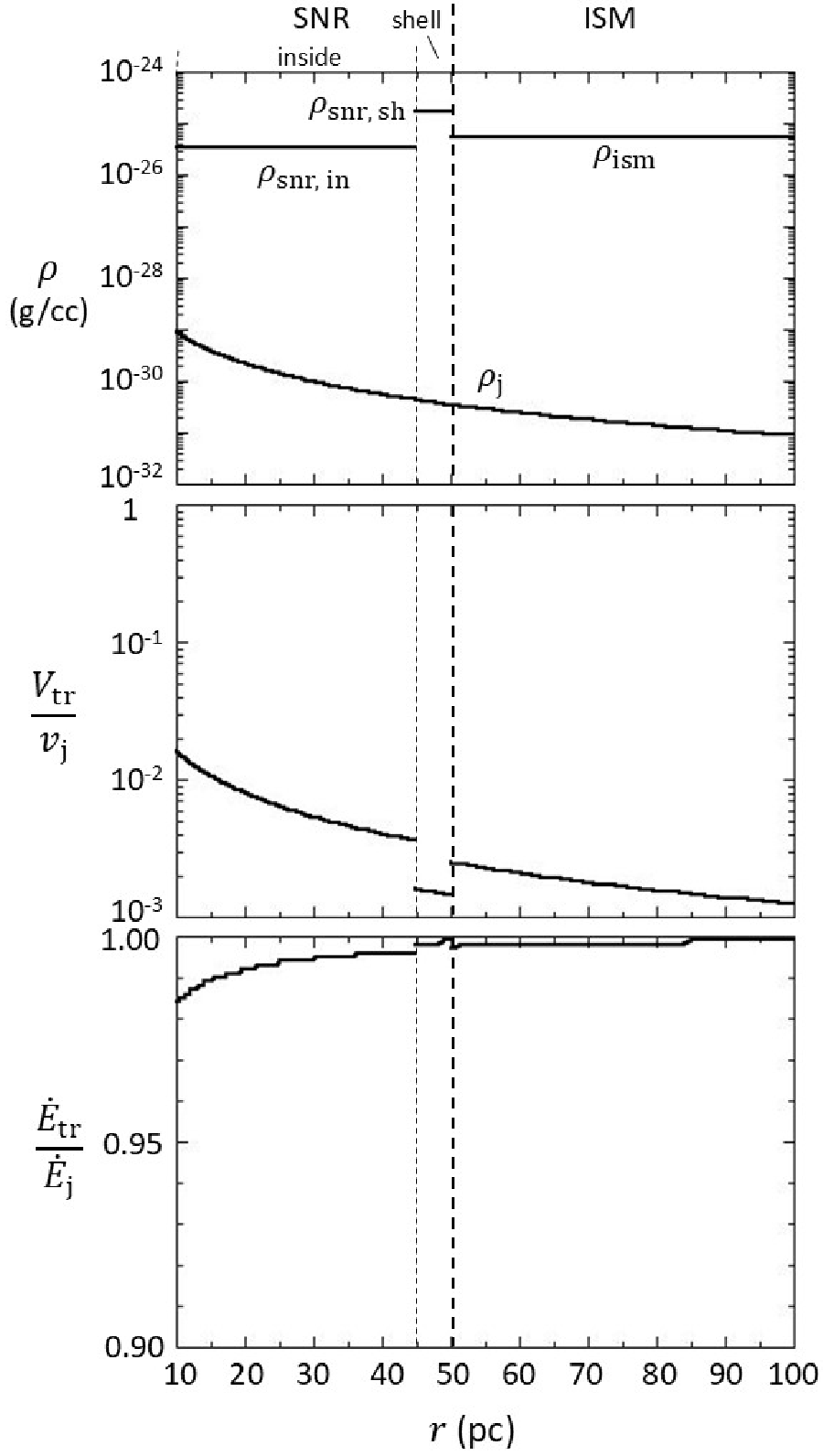} 
\end{center}
\caption{ Assumed density distributions of the SNR, the ISM and the jet (top), calculated velocity of the TR relative to the intrinsic jet velocity (middle) and calculated energy outflow rate from the TR relative to the intrinsic energy flow rate through the jet (bottom). {Alt text: Two line graphs. x axis shows distance of a jet element from 10 pc to 100 pc commonly to the three panels. y axis shows densities of the jet and surrounding matters from 10$^{-32}$ to 10$^{-24}$ g/cc in the top panel, the relative velocity from 10$^{-3}$ to 1 in the middle panel and the relative energy loss rate from 0.9 to 1 in the bottom panel, } } 
\label{fig:DensityDist}
\end{figure}

Using the density distributions of the jet and the surrounding matter as given above, 
the parameter $A$ defined in equation (\ref{eqn:A}) can be calculated as a function of $r$, which enable us to see the speed of the TR from $x$ (ratio of $V_{\rm tr}$ to $v_{\rm j}$) as a function of $A$ in equation (\ref{eqn:Root-Eq}), and the energy outflow rate from the TR derived in equation (\ref{eqn:Edot_tr}).
The results for $\dot{M}_{\rm j} = 10^{19}$ g s$^{-1}$, $\theta_{\rm j} = 1^{\circ}$ and $\theta_{\rm p} = 20^{\circ}$ are presented in figure \ref{fig:DensityDist}.
As seen from this figure, the densities of the SNR and ISM are both much larger than the jet density in the 10 - 100 pc range.
Hence, the advancing speed of the TR is largely reduced from the intrinsic jet velocity and the energy rate flowing out from the side of the TR becomes almost 100 \% of the intrinsic energy flow rate of the jet, 

\begin{figure}
\begin{center}
\includegraphics[width=8cm]{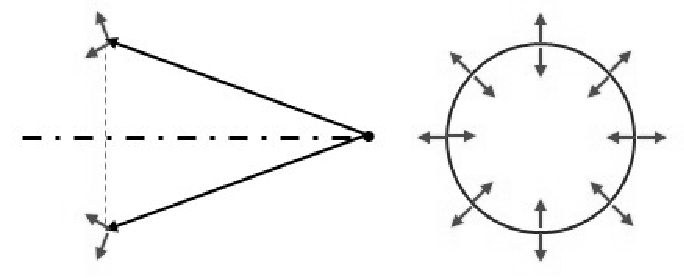} 
\end{center}
\caption{ Schematic pictures showing outflow-directions from the TR of the precessing jet for a time-interval over one precession-period. Cross sections along (left) and perpendicular to (right) the precessing axis are presented. {Alt text: Pictures show the peculiar situation of the energy flows.} } 
\label{fig:EnergyFlow}
\end{figure}

In case of a jet simply running along a straight line, the energy flow from the side of the TR could form a bow shock with an umbrella-like shape around the jet axis propagating in the surrounding medium in association with the TR advancing.
However, the advancing direction of jet in the SS433/W50 system has a precession, and thus such a situation for the directions of the energy flow as schematically shown in figure \ref{fig:EnergyFlow} could be expected.
A particular thing is that the mass- and energy-flows toward the inside of the precession cone from TRs at different phases in one precession period are all directed to the precession axis on the plane perpendicular to the precession axis. This could induce compression of the matter in the precession cone toward the precession axis and accumulation of the energy flowing out from the TR in the compressed matter around the precession axis.

\subsection{Compressed regions along the precession axis}
\begin{figure}
\begin{center}
\includegraphics[width=8cm]{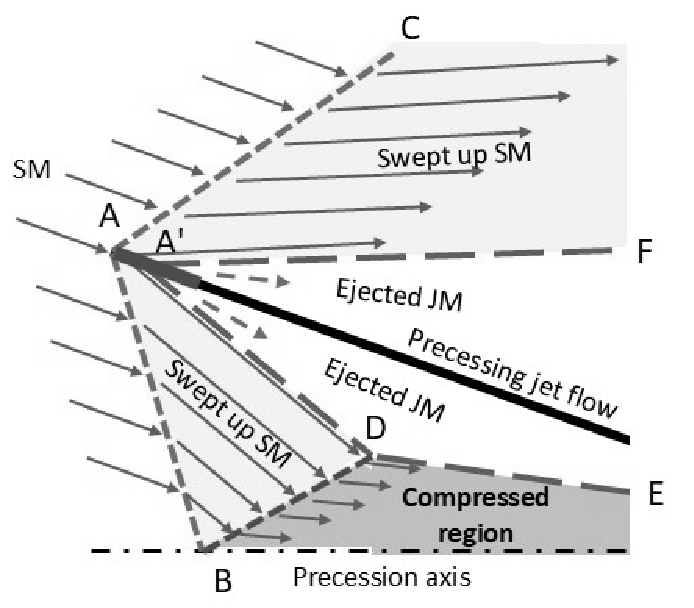} 
\end{center}
\caption{ Schematic diagram of stream lines of SM across a bow shock formed by interaction of the SM and a precessing jet on a meridian plane around the precession axis in the bow-shock rest frame. Short-dashed lines, A to B and A to C indicate the bow shock. Long-dashed lines, A' - F, and A' - D - E express the contact discontinuity between the swept up SM regions and the ejected JM regions. The SM flowing into on the inner side of the precession cone is considered to form an oblique bounce shock (dashed line between B and D) around the precession axis and eventually accumulate in the compressed region on the down-stream side of the bounce shock. {Alt text: Picture shows a process to form the compressed region.} } 
\label{fig:MatterCompression}
\end{figure}

Figure \ref{fig:MatterCompression} exhibits a schematic diagram of stream lines of SM across a bow shock formed by interaction of the SM and a precessing jet on a meridian plane around the precession axis, indicating a scenario on how a compressed region is formed around the jet precession axis. 
The jet collides with the SM and the TR appears at the top, where the SM and JM inflowing from the front shock and the rear shock respectively are ejected from the lateral side. The SM ejected from the TR successively collides with ambient SM, which forms the bow shock (short-dashed lines, A to B and A to C). The SM which has passed by the bow shock flows in directions oblique to both the bow-shock and the precessing cone, and occupy two swept up SM regions, one on the outer side of the precession cone and the other on the inner side of it. These regions sandwich two regions filled with JM ejected from the TR, again one on the outer side and the other on the inner side of the precession cone. The boundary surfaces of contact discontinuity between the SM and JM regions are indicated with long-dashed lines (A'-F and A'-D-E) in the figure. 

Since the SM flowing into the inner side of the precession cone has a velocity component directed to the precession axis as in figure \ref{fig:EnergyFlow}, the matters from different azimuthal angles should collide with one another around the precession axis, which bears an oblique bounce shock there.
Then, the SM entering the inner side of the precession cone is expected to accumulate on the down-stream side of the bounce shock, forming a compressed region.

Observationally, we know that radio lobes extend on both the east and the west side of the jet-precession axis in the SS433-W50 system.
These lobes could be the results of the matter condensation along the jet precessing axis as discussed above.
In that case, the top region of the lobe could be understood to correspond to the high density region behind the bounce shock. Indeed, the bright filamentary structure in radio (\cite{Dubner98}) and ring like features in X-rays (\cite{Brinkmann07}) are observed there in the eastern lobe.
Then, it is suggested that the TR of the jets has already passed over the SNR space and reached deep ISM space with the distance of $\sim$ 100 pc from the system center.
If so, the compressed region behind the bounce shock could have a sequence of zones of matter originally in the SNR-inside, the SNR-shell and the ISM from inside to outside, and the brightest region observed in the X-ray band could be the zone composed of the densest matter in the SNR-shell.

These predicted features of the compressed region look roughly reproducible of radio and X-ray images of the eastern lobes in the SS433-W50 system.
To see the reproducibility semi-quantitatively, we approximately estimate physical properties of the three compressed zones in table 1.
Here, we assume that the three, SNR-inside, SNR-shell and ISM regions in the precession cone before the bow shock passage are simply compressed to the three respective zones after the passage.
$V'$ on the first line is each volume of the three compressed zones,  Each of the compressed volume is assumed to be a fourth of the respective original volume, since the opening angle of the lobes is about a half of that of the precession cone. $n'$ on the second line is the average number density of the compressed zone.
$\Delta E$ on the third line is the amount of energy flowing out from the TR in the period during which the TR has passed over the respective region. If the region has a radial range from $r_{1}$ to $r_{2}$, it is calculated as
\begin{equation}
\Delta E = \int_{r_{1}}^{r_{2}} \dot{E}_{\rm tr} \frac{dr}{V_{\rm tr}}.
\label{eqn:DE}
\end{equation}
The values obtained with $\dot{E}_{\rm tr}$, $V_{\rm tr}$ and the range of $r$ shown in figure \ref{fig:DensityDist} for the three regions are presented in table 1.
$u$ on the fourth line is the average energy density of the compressed zone calculated as
\begin{equation}
u = \frac{\Delta E/2}{V'},
\label{eqn:u}
\end{equation}
where the reason why $\Delta E$ is divided by 2 in the numerator is because a half of the outflowing energy from the TR goes to the inner side of the precession cone.

\begin{table}
\caption{Expected properties of the compressed regions}
\begin{center}
\begin{tabular}{c|ccc}
\hline
Original regions & SNR-inside & SNR-shell & ISM \\ 
\hline
$V'$ (cm$^{3}$) \footnotemark[(1)] & $8.5 \times 10^{58}$ & $3.2 \times 10^{58}$ & $8.2 \times 10^{59}$ \\
$n'$ (cm$^{-3}$) \footnotemark[(2)] & $8 \times 10^{-2}$ & $4 \times 10^{-1}$ & $1.3 \times 10^{-1}$ \\
$\Delta E$ (erg) \footnotemark[(3)] & $1.8 \times 10^{50}$ & $1.5 \times 10^{50}$ & $9.0 \times 10^{50}$ \\
$u$ (erg cm$^{-3}$) \footnotemark[(4)] & $2.1 \times 10^{-9}$ & $3.3 \times 10^{-9}$ & $1.1 \times 10^{-9}$ \\
\hline
\end{tabular}
\end{center}
\label{Table1}
\begin{tabnote}
\footnotemark[$(1)$] Compressed volume \\ 
\footnotemark[$(2)$] Compressed number density\\
\footnotemark[$(3)$] Energy input from the TR of the jet \\
\footnotemark[$(4)$] Average energy density
\end{tabnote}
\end{table}

The energy density $u$ could be divided into densities of the bulk kinetic energy, $u_{\rm bk}$, the thermal energy, $u_{\rm th}$, the magnetic energy, $u_{\rm mg}$ and the non-thermal energy, $u_{\rm nt}$.

Since the SM flowing into the TR is accelerated to the velocity, $V_{\rm tr}$, in the TR advancing direction and flows out with that forwarding velocity from the side of the TR, the momentum outflowing rate is roughly estimated to be $\rho_{\rm sm} V_{\rm tr} 4\pi r^{2} \theta_{\rm pc} \theta_{\rm j} V_{\rm tr}$.
A half ot this outflowing momentum should be transferred to the SM matter inflowing to the bow shock front within the precession cone and thus we can have the following equation as
\begin{equation}
\frac{1}{2} \rho_{\rm sm} V_{\rm tr} 4\pi r^{2} \theta_{\rm pc} \theta_{\rm j} V_{\rm tr} \simeq \rho_{\rm sm} V_{\rm tr} \pi (r \theta_{\rm pc})^{2} v',
\label{eqn:v'-Eq}
\end{equation}
where $v'$ is the bulk velocity of the compressed region and $\theta_{\rm pc}$ is the half openning angle of the precession cone.
From this equation we get
\begin{equation}
v' \simeq 2 \left( \frac{\theta_{\rm j}}{\theta_{\rm pc}}\right) V_{\rm tr} \simeq \frac{1}{10} V_{\rm tr}
\label{eqn:v'-Cal}
\end{equation}
for $\theta_{\rm j} = 1^{\circ}$ and $\theta_{\rm pc} = 20^{\circ}$.
Thus, if we assume that the density of the matter in the compressed region, $\rho_{\rm cr}$, is 4 times $\rho_{\rm sm}$, the bulk kinetic energy density of the compressed region is approximately given as 
\begin{equation}
u_{\rm bk} \simeq \rho_{\rm cr} \frac{v'^{2}}{2} \simeq 4 \rho_{\rm sm} \frac{1}{100} \frac{V_{\rm tr}^{2}}{2}
\label{eqn:u_bk-Cal}
\end{equation}
and is estimated to be much less than 10$^{-11}$ erg cm$^{-3}$ for the $\rho_{\rm sm}$ and $V_{\rm tr}$ values in figure \ref{fig:DensityDist}, the contribution of which to $u$ in table \ref{Table1} is negligibly small.
The fraction of the non-thermal energy density, $u_{\rm nt}$ to the total energy density, $u$, is also negligibly small as discussed later.
Then, if we consider that the total energy density is equipartitioned to the remaining two components, $u_{\rm th}$ and $u_{\rm mg}$, they are given as
\begin{equation}
u_{\rm th} = u_{\rm mg} =\frac{u}{2}.
\label{eqn:u_th-u_mg}
\end{equation}

\subsubsection{Predicted properties of thermal emissions}
From the given $u_{\rm th}$, we can estimate the average temperature of the compressed matter, $T'$ via the equation, $u_{\rm th} \simeq 3 n' k T'$.
Then, we can further calculate luminosities of thermal emission from the compressed zones, $L_{\rm th}$, with an equation as
\begin{equation}
L_{\rm th} \simeq n'^{2} V' \Lambda (T'),
\label{eqn:L_th-Cal}
\end{equation}
where $\Lambda(T)$ is the cooling function for a thermal plasma, 
Table \ref{Table2} exhibits the calculated values of $u_{\rm th}$, $kT'$ and $L_{\rm th}$ for the three compressed zones.
In the $L_{\rm th}$ estimation, we have adopted $\Lambda = 2 \times 10^{-23}$ erg cm$^{3}$ s$^{-1}$ which is the value around $kT \sim$ 0.5 - 1 keV in case of the solar metallicities (see Fig. 8 in \cite{Sutherland93}).

\begin{table}
\caption{Expected thermal emissions from the compressed zones}
\begin{center}
\begin{tabular}{c|ccc}
\hline
Original regions & SNR-inside & SNR-shell & ISM \\ 
\hline
$u_{\rm th}$ (erg cm$^{-3}$) & $1.1 \times 10^{-9}$ & $1.7 \times 10^{-9}$ & $5.5 \times 10^{-10}$ \\
$kT'$ (keV) & 2.3 & 0.9 & 0.9 \\
$L_{\rm th}$ (erg s$^{-1}$) & $1.1 \times 10^{34}$ & $1.0 \times 10^{35}$ & $2.9 \times 10^{35}$ \\
\hline
\end{tabular}
\end{center}
\label{Table2}
\end{table}

The X-ray bright region in the eastern lobe could correspond to the SNR-shell zone, where the observed X-ray luminosity of the thermal component is $\sim 4 \times 10^{34}$ erg s$^{-1}$ for the source distance of 5.5 kpc and $kT$ obtained by the spectral analysis is $\sim$0.2 keV (\cite{Brinkmann07}).
On the other hand, the very eastern region having the radio bright structure often called as ``ear" or ``terminal shock" could corresponds to the ISM zone, where the observed X-ray luminosity of the thermal component is $\sim 2 \times 10^{35}$ erg s$^{-1}$ and the estimated $kT$ is $\sim$0.2 keV (\cite{Brinkmann96}).
These values agree to respectively corresponding values in table \ref{Table2} within a factor of 3 $\sim$ 4.

\subsubsection{Predicted properties of synchrotron emissions}
From the given $u_{\rm mg}$, we can estimate the average magnetic fields for the three zones via the equation, $u_{\rm mg} = B^{2}/8\pi$.
Then, if non-thermal electrons exist there, synchrotron emission is expected.

We suppose that the number density of the electrons, $n_{\rm e}$, has a distribution on the Lorentz factor, $\gamma$ as
$n_{\rm e} (\gamma) = \frac{C}{ m_{\rm e} c^{2}}u_{\rm e} \gamma^{-p}$ in the range between $\gamma_{\rm min}$ and $\gamma_{\rm max}$, where $u_{\rm e}$ is the energy density of the non-thermal electrons and $C$ is given as
\begin{equation}
C = \begin{cases}
(p-2)\gamma_{\rm min}^{p-2} & \text{when $p > 2$} \\
\left[ \ln \left( \frac{\gamma_{\rm max}}{\gamma_{\rm min}}\right) \right]^{-1} & \text{when $p = 2$} \\
(2-p) \gamma_{\rm max}^{p-2} & \text{when $p < 2$},
\end{cases}
\label{eqn:C}
\end{equation}
for $\gamma_{\rm max} \gg \gamma_{\rm min}$.
$m_{\rm e}$ and $c$ are the electron mass and the light velocity respectively.
Then, the synchrotron luminosity around a photon frequency, $\nu$, is given as (see \cite{Rybicki79})
\begin{eqnarray}
\nu L_{\rm sy, \nu} &\simeq& \frac{3^{(p+2)/2}}{2\pi (p+1)} \Gamma \left (\frac{p}{4}+\frac{19}{12}\right) \Gamma \left(\frac{p}{4} - \frac{1}{12}\right) C \nonumber \frac{\sigma_{\rm T}}{m_{\rm e} c} \\ && V' u_{\rm e} u_{\rm mg} (\sin \phi)^{(p+1)/2} \left(\frac{\nu}{\nu_{\rm B}}\right)^{(3-p)/2},
\label{eqn:nuLnu}
\end{eqnarray}
where $L_{\rm sy, \nu}$ is the synchrotron luminosity per frequency, $\Gamma(x)$ is the gamma function of $x$, $\sigma_{\rm T}$ is the Thomson cross section, $\phi$ is a pitch angle of gyrating electron, and $\nu_{\rm B} \equiv eB/(2 \pi m_{\rm e} c)$. 

We have the $\nu L_{\rm sy, \nu}$ values from observations of the eastern lobe in X-rays and radio, which are indicated for the radio bright, top region (considered to be the ISM zone in the compressed region) and the X-ray bright, mid region (to be the SNR-shell zone) in table \ref{Table3}.
In the radio band, the top region is bright but the mid region has no particular brightness-enhancement from the main spherical region of W50 (see \cite{Dubner98}). 
The $\nu L_{\rm sy, \nu}$ value at $\nu_{\rm R}$ = 1.4 GHz of the top region is roughly calculated from the observed brightness per steradian of \citet{Dubner98}, while a tenth of the value for the top side is adopted as the upper limit to the value of the mid region. 
On the other hand, in the X-ray band, the non-thermal emission is bright in the mid region but dim in the top region.
The $\nu L_{\rm sy, \nu}$ value of the non-thermal component around $\nu_{\rm X} = 2.4 \times 10^{14}$ Hz ( 1 keV) from the mid region is obtained from \citet{Brinkmann07} and a tenth of this value is adopted as the upper limit to the value of the top region.

Now, we can calculate the spectral index, $\alpha$, of each of the synchrotron emissions from the two regions by using the following equation as
\begin{equation}
\alpha = 1-\log \frac{(\nu L_{\rm sy, \nu})_{\rm X}}{(\nu L_{\rm sy, \nu})_{\rm R}} / \log \frac{\nu_{\rm X}}{\nu_{\rm R}}.
\label{eqn:alpha-Cal}
\end{equation}
The obtained ranges of $\alpha$ for the two regions are presented in table \ref{Table3}. 
These ranges seem consistent with the results obtained from radio observation itself by \citet{Dubner98}.
If we assume the $p$ values in the ranges calculated with the relation of $p = 2 \alpha + 1$ from the $\alpha$ ranges for the two regions as in table \ref{Table3}, we can estimate $u_{\rm e}$ from equation (\ref{eqn:nuLnu}) using the values of $V'$, $u_{\rm mg}$, $B$ in the respective zones and the values of 
$(\nu L_{\rm sy, \nu})_{\rm R}$ at $\nu_{\rm R}$ and $(\nu L_{\rm sy, \nu})_{\rm X}$ at $\nu_{\rm X}$ for the top region and the mid region respectively, and approximating $\sin \phi =1$. In calculations of the $C$ values with equation (\ref{eqn:C}), we have assumed $\gamma_{\rm max}/\gamma_{\rm min} = 10^{10}$ for $p=2$ and $\gamma_{\rm min} = 1$ for $p=2.6$.  The results are shown in the table in the form of a ratio of $u_{\rm e}$ to $u_{\rm mg}$,

\begin{table}
\caption{Properties of the synchrotron emissions}
\begin{center}
\begin{tabular}{c|cc}
\hline
Regions in & Mid region & Top region \\ 
the eastern lobe & (SNR-shell zone) & (ISM zone) \\
\hline
$u_{\rm mg}$ (erg cm$^{-3}$) & $1.7 \times 10^{-9}$ & $5.5 \times 10^{-10}$ \\ 
$B$ (gauss) & $2.0 \times 10^{-4}$ & $1.2 \times 10^{-4} $ \\
$(\nu L_{\rm sy, \nu})_{\rm R}$ (erg s$^{-1}$) & $\lesssim 2.3 \times 10^{30}$ & $2.3 \times 10^{31}$ \\
$(\nu L_{\rm sy, \nu})_{\rm X}$ (erg s$^{-1}$)& $2.4 \times 10^{34}$ & $\lesssim 2.4 \times 10^{33}$ \\
$\alpha$ & $\lesssim 0.51$ & $\gtrsim 0.76$ \\
$p$ (assumed) & 2 & 2.6 \\
$u_{\rm e}/u_{\rm mg}$ & $2.4 \times 10^{-5}$ & $3.1 \times 10^{-4}$ \\
\hline
\end{tabular}
\end{center}
\label{Table3}
\end{table}

\section{Summary and discussion}\label{sec:4}
We have first investigated the approximate structure of the TR (top region) of a jet where the jet matter interacts with the surrounding matter in section \ref{sec:2}.
The schematic cross section of the TR considered here is shown in figure \ref{fig:TR}.
In this picture, the jet matter inflowing from the rear shock collides with the surrounding matter inflowing from the front shock at the contact discontinuity surface and the strengthened pressure at the surface pushes both the matters out from the side of the TR as viewed in the TR rest frame.
Approximate calculations reveal that the advancing speed of the TR and the energy flow rate carried by the outflowing matters from the side are determined as a function of the ratio of the jet matter density, $\rho_{\rm j}$, to the surrounding matter density, $\rho_{\rm sm}$.

When $\rho_{\rm j} \gg \rho_{\rm sm}$, the deceleration of the jet due to the interaction with the SM is very small and the energy outflowing rate from the TR is also very small.
When $\rho_{\rm j} \ll \rho_{\rm sm}$, on the other hand, the TR is significantly decelerated by the interaction with the SM and almost all of the energy flow rate through the jet is converted the energy flow rate from the lateral side of the TR to the surrounding space.

Next, the model considerations have been applied to jet - surroundings interactions in the SS433 - W50 system in which the jet ejection direction from the central compact object precesses with the precession angle of 20$^{\circ}$.
The eastern and western lobes extend in directions opposite to each other, along the precession axis, indicating that the elongated structures could be results of interaction between the jets and the surroundings.
However, the half opening angle of the lobes is about 10$^{\circ}$ and is significantly smaller than the precession angle of 20$^{\circ}$.

Although one possibility to explain this angle difference between the inner region and the outer region could be that the jet-advancing direction has been changing with distance by a focusing of the hollow cone trajectory or that the precession angle has been evolving with time to the present value, we have adopted another situation here in which the advancing direction of the jets is kept constant from the central engine to the present terminal position on the surface of the same precession cone.
In this situation, the surrounding matter trapped within the precession cone is discussed to be compressed around the precession axis by the energy outflow from the TR side, in section \ref{sec:3}.
The region around the precession axis in which the matter within the precession cone is compressed to the structure with a significantly smaller half-cone angle than 20$^{\circ}$ is expected to be the most luminous region in the present scenario.

Emissions from the TR itself are considered to be negligibly weak compared to those from the compressed region
since the mass and the volume in the TR are much smaller than those of the compressed region, .
On the other hand, the same amount of energy, $\Delta E$, as in table \ref{Table1} outflows also to the outer side of the precession cone after the passage of the TR over the respective region. However, the energy could dilute in the larger space as the bow shock front advances and it could be impossible for the outer side to exhibit such enhanced emissions as from the compressed region on the inner side.

The compressed region is considered to be lengthening along the precession axis in association with the advance of the jet-TR by taking anterior matter in through the bow shock and the bounce shock as schematically shown in figure \ref{fig:MatterCompression} and to have a radial density distribution which reflects that of the surrounding matter before the jet passage.
Based on this consideration, we have inferred that the ``ear" structure observed near the eastern end of the radio lobe could correspond to the backside of the bounce shock and that the X-ray bright knot observed in the mid of the eastern lobe could be the place where the relatively dense matter in the SNR shell was compressed. 

Three bright regions are observed from the eastern lobe in X-rays named ``e1", ``e2" and ``e3" in distance order (\cite{SafiHarb97}). The e3 position agrees to the ``ear" in radio and the X-ray brightest region e2 could be produced by the matter on the backside of the SNR front shock. The e1 region might consist of the matter on the backside of the SNR reverse shock.

It is interesting to note that wavy and/or filamentary structures are seen along the edge of the X-ray brightest region (see e.g. Fig.7 of \cite{Brinkmann96}). In the present scenario, it is considered that the X-ray brightest region exists in the compressed region of the swept up SM and that the edge is the contact discontinuity with the ejected JM (cocoon) region. Numerical simulations tell us that the contact discontinuity often gets unstable due to the Kelvin-Helmholtz instability or the Rayleigh-Taylor instability and has wavy or filamentary surface (e.g. \cite{Falle91}; \cite{Ohmura21}). The structures seen at the edge of the X-ray brightest region in the eastern lobe could represent the unstable situation of the contact discontinuity.
We can also extend a similar conjecture to a structure called ``chimney" observed at the north-east end of the eastern lobe in radio (\cite{Dubner98}). The position of this structure can be regarded to exist along the contact continuity A'-D in figure \ref{fig:MatterCompression} in the present scenario. Hence, the ``chimney" might be a result of unstable situation of the contact discontinuity.

Semi-quantitative estimations of physical quantities have been done to confirm the above inference on the origin of the eastern lobe.
The energies flowing out from the TR after the passage over the SNR-inside, SNR-shell and ISM regions are respectively calculated with the model in section \ref{sec:2} and the respective energy densities are obtained assuming that the respective volumes within the precession cone are compressed to a forth of the original ones. This compression factor is due to the observation that the opening angle of the eastern lobes is about a half of that of the precession cone. 
The results are tabulated in table \ref{Table1}. 

Supposing equipartition between the thermal and magnetic energy densities, properties of thermal emissions from the three zones have been derived in table \ref{Table2}. The results on the SNR-shell zone and ISM zone roughly reproduce the observed values of the mid, X-ray bright region and the top, radio bright region of the eastern lobe respectively.
These results depend on the assumptions of the density values of the surrounding matter and the jet.
The luminosity of the thermal emission calculated with equation (\ref{eqn:L_th-Cal}) depends largely on the density of the surrounding matter, since the temperature dependence of $\Lambda$ is weak in the temperature range around several 0.1 keV. 
Considering that the calculated luminosities with the present densities roughly match the observed values, there could be no large room to deviate the density values of the surroundings from the premised values in figure \ref{fig:DensityDist}, as far as the compression factor of the compressed region is fixed to be $\sim$4.
On the other hand, $kT$ is determined by $u/n'$ and $u$ is proportional to $\Delta E/V'$.
Thus, $kT$ is proportional to $\Delta E$ when $n'$ and $V'$ are given.
$\Delta E$ is obtained through equation (\ref{eqn:DE}), and is $\propto \dot{E}_{\rm tr}/V_{\rm tr}$. Since $\dot{E}_{\rm tr} \simeq \dot{E}_{\rm j} \propto \dot{M}_{\rm j}$ and $V_{\rm j} \propto \sqrt{A} \propto \sqrt{\dot{M}_{\rm j}}$ via equation (\ref{eqn:Root-Eq}) when $A \ll 1$, we finally get $kT \propto \sqrt{\dot{M}_{\rm j}}$.
The $kT$ values of SNR-shell and ISM calculated for $\dot{M}_{\rm j} = 10^{19}$ g s$^{-1}$ in table \ref{Table2} are slightly larger than the observed values. The smaller $\dot{M}_{\rm j}$ value than the usually adopted value of 10$^{19}$ g s$^{-1}$ is preferable at least from this viewpoint.

For the synchrotron emission from the top radio bright region and the mid X-ray bright region of the eastern lobe, we have first derived the spectral slope from the radio band to the X-ray band from observations approximately and then estimated the energy density of the non-thermal electrons with the help of the magnetic energy density calculated on the approximation of the equipartition with the thermal energy.
The results are tabulated on table \ref{Table3}.
The value of $u_{\rm e}/u_{\rm mg}$ in the top region is larger than that in the mid region by an order of magnitude. 
This could be reasonable, if we consider that the top region corresponds to the backside of the bounce shock where the particle acceleration is just in progress, while the mid region is a place where the particle acceleration ceased and the accelerated particles have been diluted.
The $u_{\rm e}/u_{\rm mg}$ value of several times 10$^{-4}$ could also be understandable, if several \% of the thermal energy density is spent to the particle acceleration and about 1 \% of the accelerated particle energy is for electrons.

On the other hand, the slope of the synchrotron emission from the top region is larger than that from the mid region. This might be due to the following situation: 
In the top region, accelerated electrons are originated from fresh thermal electrons of the inflowing ISM matter, while accelerated electrons in the mid region significantly include those from non-thermal electrons which have once accelerated at the backside of the front shock or the reverse shock in the SNR. The double accelerations could make the spectrum flatter.
The situation of the repeated accelerations through the SNR shock, the bow shock and the bounce shock could give an opportunity for electrons to get very high energy, which could explain why 10 - 100 TeV gamma-rays are detected from the X-ray brightest regions in the eastern and western lobes (\cite{Abeysekara18}; \cite{HESS24}).

According to the present scenario, it is inferred that the TR has already reached to a place with the distance of $R_{\rm tr, t} \sim$ 100 pc from the central compact object. The required time, $t_{\rm j}$, for the TR to travel over such a distance can be estimated by integrating $dr/V_{\rm tr}$ from the initial position to the present TR position 
and is calculated to be of the order of $10^{5}$ yr for the parameters in figure \ref{fig:DensityDist}.
According to the Sedov solution, we have the following relation as
\begin{equation}
R_{\rm snr} \simeq 5 \times 10 \ t_{\rm snr, 5}^{2/5} E_{\rm snr, 51}^{1/5} n_{\rm ism, -1}^{-1/5} \mbox{ pc},
\label{eqn:R_snr-Sedov}
\end{equation}
where $t_{\rm snr}$ is the age of the SNR in unit of 10$^{5}$ yr, $E_{\rm snr, 51}$ is the total energy of the SNR in unit of 10$^{51}$ erg and $n_{\rm ism, -1}$ is the number density of the interstellar matter surrounding the SNR in unit of 10$^{-1}$ cm$^{-3}$.
The $R_{\rm snr}$ value is roughly consistent with the present situation of the SNR in W50, when all the parameters in this equation (\ref{eqn:R_snr-Sedov}) are around unity. 
The Sedov solution also tells us the temperature behind the shock front, $T_{\rm snr}$, as 
\begin{equation}
T_{\rm snr} \simeq 5 \times 10^{5} \ t_{\rm snr, 5}^{-6/5} E_{\rm snr, 51}^{2/5} n_{\rm ism, -1}^{-2/5} \mbox{ K}.
\label{eqn:T_snr-Sedov}
\end{equation}
If we calculate the radiative cooling time for $T_{\rm snr} = 5 \times 10^{5}$ K and $n_{\rm snr, sh} = 0.1$ cm$^{-3}$ adopting $\Lambda(T_{\rm snr}) = 2 \times 10^{-22}$ erg cm$^{3}$ s$^{-1}$ from \citet{Sutherland93}, it is $3 \times 10^{5}$ yr.
The SNR in W50 could now be in a transition from the Sedov phase to the cooling phase.
The age of the SNR is also argued to be $\sim 10^{5}$ yr from a viewpoint of motion of the SS433 system in relation to the Galactic plane by \citet{Dubner98}.
It should be noted here that the total energy transferred from the jets to the SNR is approximately estimated by summing the $\Delta E$ values for the SNR-inside and SNR-shell in table \ref{Table1} and multiplying 2 for the two opposite jets to be close to 10$^{51}$ erg s$^{-1}$, the typical total energy of s SNR. This energy transfer could give a significant effect on the evolution of the SNR but we don't discuss it further here.

As shown above, the present scenario on the interaction of the jets with the surroundings in the distance range of 10 - 100 pc can explain the overall properties of the SS433-W50 system in that scale.
Inoue (2022) proposed a scenario to interpret the overall observed features of the SS433-W50 system, in which a basic idea on the extended lobes was presented but the considerations were insufficient.
This paper revises that part (subsection 3.7 and figure 7) of Inoue (2022).

According to discussions in \citet{Inoue22} and the present work, an element of the jet matter experiences interactions with the ambient matter three times at different distances with different observational appearances from one another.

The first and second ones are interactions with the disk wind matter, which were discussed by \citet{Inoue22}.
The disk wind is considered to flow outward from an accretion ring (\cite{Inoue21}).
An accretion disk also extends inward from the accretion ring and ejects jets from the innermost side (see \cite{Inoue24} for a possible jet-ejection mechanism).
Supposing that the accretion ring is excited to precess (\cite{Inoue12}; \cite{Inoue19}), both the jets and the disk wind come to precess linked with the accretion ring motion.
Then, all the jet matter ejected from the central engine is considered to encounter with disk wind matter first around a distance of $v_{\rm dw} t_{\rm p}/2 \simeq 7 \times 10^{14}$ cm for the wind velocity $v_{\rm w} \simeq 10^{8}$ cm s$^{-1}$, if the opening angle of the wind from the equatorial plane is larger than 50$^{\circ}$.
This is the first interaction of the jet matter and the disk wind matter, which could cause the enhancement of the optical lines and the following radio brightening of the jets (e.g. \cite{Fabrika04}). Kicking away of the disk wind matter from the jet path could also happen as the result of this interaction.
Even after this first encounter, the jet goes through the disk wind envelope every distance of $v_{\rm w} t_{\rm p}$ but nothing happens since the disk wind matter on the jet path has already been evacuated at the first interaction with the jet.
However, interaction of the disk wind with the SNR matter is inferred to press and stir the disk wind matter so that the vacant region along the jet path in the disk wind envelope could be refilled with the ambient disk wind matter at a distance of $\sim 10^{17}$ cm and the refilled disk wind matter could collide with the jet matter again.
This is the second interaction of the jet matter with the disk wind matter, which could reheat the jet matter as observed in X-rays (\cite{Migliari02}).
The distances of the first and second interaction places are always the same, where the disk wind steadily supplies the matter kicked out by the jet matter.
Since the density of the disk wind is much less than that of the jet at the same distance, the deceleration of the jet speed should be negligibly small in the first and second interactions with the disk wind.

The last one is the interaction with the SNR matter or the ISM, which has been studied in this paper.
It happens at the top of the jet and all the jet matter finishes its journey through the jet track there by colliding with the SNR matter or the ISM and flowing out from the lateral side.
The top region advances by being pushed by newly coming jet matter and pushing out the anterior SNR matter or ISM successively, elongating the evacuated space of the surrounding matter along the jet-track and enabling the following jet matter to go further.


\bibliographystyle{plain}
\bibliography{SS433-References}

\end{document}